\documentclass[aip,apl,amsmath,amssymb,reprint]{revtex4-1}
\usepackage[dvipdfmx]{graphicx}
\usepackage{physics}

\begin{document}
\title{Single time pixel imaging enabled by repurposing optoelectronic devices}

\author{Ryota Keyaki}
\affiliation{Graduate School of Arts and Sciences, University of Tokyo, Komaba, Meguro, Tokyo 153-8902, Japan}

\author{Susumu Fukatsu}
\email{sjfkatz@g.ecc.u-tokyo.ac.jp}
\affiliation{Graduate School of Arts and Sciences, University of Tokyo, Komaba, Meguro, Tokyo 153-8902, Japan}

\begin{abstract}
One-time readout temporal ghost imaging is attempted by utilizing optoelectronic devices that are not originally intended for signal photon detection purposes and as such slow by design. 
A visible light-emitting diode having a response time $\tau$=0.036\,ms and a solar cell with $\tau$=\,3.1\,ms are used to retrieve a rectangular pulse train, which is otherwise rounded with significant overlapping, in the image of a temporal mask simply by capturing data once at a selected single time pixel followed by division. Appropriate quality metrics and effective operation duration are discussed.
\end{abstract}

\maketitle
\newpage
Ghost imaging in the time domain is an enabling technology\cite{ryczkowski2016ghost,shirai2010temporal, chen2013temporal, devaux2016computational, devaux2016temporal, ryczkowski2017magnified, o-oka2017,ryczkowski2017magnified,liu2018high,tang2018computational,wu2019temporal,haojie2019research,huang2019stable,huang2020temporal,wu2020temporal}. It allows retrieval of images of a time-varying object based on the information about a bipartite sharing very little in common. 
Also known as temporal ghost imaging (TGI), it arguably finds a rich variety of applications including metrology, telecommunication, and information/signal processing\cite{hualong2020information,tian2020acoustic,wenwen2020fourier,wu2023general}.
Because of its inherent two-arm geometry,  the way to detect light can make a difference as in the case of the conventional GI pertaining to the space domain\cite{pittman1995optical,karmakar2012ghost,shih2012physics,moreau2018ghost}. 
Single pixel imaging (SPI) is such an example as regards GI\cite{edgar2019principles,zhao2021compressive}. 
With photons, if not all, concentrated onto a single-pixel detector, SPI literally allows imaging of an object without resolving it nor even directly seeing it.  
Moreover, SPI offers a diverse set of potential uses\cite{edgar2019principles}, not just imaging.

Meanwhile, a time-domain analog of SPI is emerging\cite{tang2018single,Y_K_Xu, J_Zhao_TDSinglePixel,Xinwei_CTGI_LED}.
Although largely unexplored, a facile while viable TGI protocol was recently proposed\cite{o-oka2023} that allows \emph{literal} single time pixel imaging (STPI). 
It is essentially a one-time detector readout followed by division. 
The readout is made only once at a selected single time pixel on the trailing edge of the detector output in a TGI configuration. 
On one hand, this compares with the previous TGI attempts to which time integration of photon counts is pivotal\cite{ryczkowski2016ghost,o-oka2017}.  
On the other hand, common optoelectronic devices of slow response can have an added advantage because of this\cite{ryczkowski2016ghost}. 
In fact, one could utilize or repurpose those devices which are not originally intended for detecting signal-encoded photons. 
Light-emitting diodes (LED) as illuminator and solar cells (SC) as energy harvester are the examples. 

In this work, we report \emph{semi}-"computational"\cite{shapiro2008computational} implementation, i.e., missing the reference arm in part,  of the STPI from the versatility perspective, as opposed to the previous fully-implemented physical STPI attempt\cite{o-oka2023}. 
Repurposing optoelectronic devices like light-emitting diode and solar cell for use as the signal-photon detector is demonstrated for a wide time frame up to 10 ms. Time-dependent quality measures of the retrieved images, and how long the STPI works effectively are discussed.

\begin{figure}[!b]
\begin{center}
\includegraphics[width =.9 \linewidth]{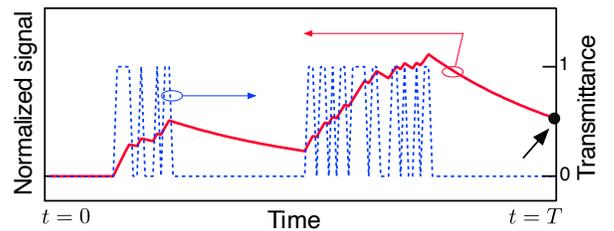}
\end{center}
\vspace{-5mm}
\caption{Temporal mask profile (broken) and slow detector output (solid) as a function of time. Single time-pixel imaging, i.e., one-time readout temporal ghost imaging, is performed by reading the slow detector output at a selected point on the leading edge after the last light input, e.g., at the end of the frame, $T$, indicated by the solid dot and arrow (black). }
\label{Fig.1}
\end{figure}

Figure \ref{Fig.1} schematically shows a typical time-evolution of the slow-detector output (solid line). Also shown by the broken line is the transmittance profile of an object, i.e., temporal mask.  Our objective is to retrieve the latter from the former. The one-time signal readout at the end of the time frame $I(T)$ is given by its convolution with the impulse response function of the detector $\mu$,
\begin{equation}
I(T)\!=\!\int^T_0\!\!dt' \mu(T\!-\!t')I(t').
\label{eq1}
\end{equation}
Our STPI, or one-time readout TGI\cite{o-oka2023}, is based on the cross-covariance $C(t)$ of joint photon detection events. 
The probability of such an event is given in terms of the normalized product of the intensity of light along the two arms. 
For a mask with a fractional transmittance, $M(t)$,  
\begin{align}
C(t)\!&=\!\ev{\Delta I_{\rm R}(t)\Delta I_{\rm T}(T)} \nonumber \\
&=\!\int^T_0\!\!\!\!dt'\mu(T-t')M(t')\ev{\Delta I_{\rm R}(t)\Delta I_{\rm T}(t')} \nonumber\\
&=\!\mu(T-t)M(t)
\label{eq2}
\end{align}
assuming an autocorrelation $\ev{\Delta I_{\rm R}(t)\Delta I_{\rm T}(t')}\!\!=\!\!\delta(t\!-\!t')$. Here $\ev{\cdot}$ stands for the ensemble average, R for the reference arm, and T for the test arm with the object. 
Equation\,(\ref{eq2}) shows that prior knowledge of $\mu(T-t)$ allows $M(t)$ to be obtained through division of $C(t)$ by $\mu$.
This is in contrast with the previous \emph{computational} attempts\cite{tang2018single,Xinwei_CTGI_LED} calling for not a little, if not massive, data handling and hence with a certain degree of complexity.

\begin{figure}[!t]
\begin{center}
\includegraphics[width =0.8 \linewidth]{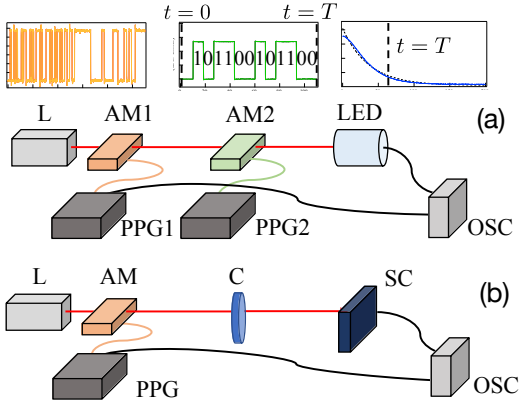}
\end{center}
\vspace{-5mm}
\caption{STPI configured with a (a) visible LED and (b) solar cell as the photodetector. L, light source; AM1,2, Acousto-optic modulator; PPG1,2, Pulse pattern generator; LED, SC, solar cell; light-emitting diode; OSC, oscilloscope. Note that beam blanking electronics was used to generate burst pulse patterns for (b). Time-based plots of AM1, AM2, and LED output are illustrated in the panels in this order from the left.}
\label{Fig.2}
\end{figure}

\begin{figure}[!b]
\begin{center}
\includegraphics[width =1 \linewidth]{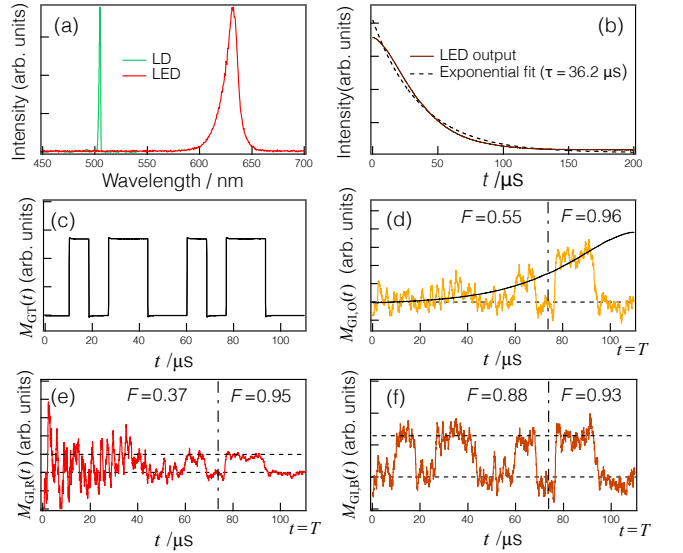}
\end{center}
\vspace{-5mm}
\caption{(a) Emission spectra of LD (green) and LED (red). (b) (Solid) Trailing-edge part of response function of LED as photodetector, $\mu_{\rm LED}$. (Broken) Exponential fit $\exp(-t/\tau)$ with $\tau=36\,\mu$s. (c) As-captured temporal mask profile. (d) As-captured STPI image (orange) and time-reversed $\mu_{\rm LED}$ (black). (e) Division-corrected STPI image. (f) TGI imgae.}
\label{Fig.3}
\end{figure}

First, our STPI was attempted using a visible LED as the photodetector\cite{Xinwei_CTGI_LED}.
Figure\,\ref{Fig.2}(a) shows the schematic setup. 
A visible LED (OptoSupply OSHR5161A-QR) designed to emit in the range 620-630 nm was used. 
In the beginning, the impulse response of the LED, $\mu_{\rm LED}$, was measured.
The light source was a 505-nm cw laser diode (LD, OxLasers A-G100F-P).
The output of the LD was on-off modulated by an acousto-optic modulator (AM, Gooch\&Housego 3080-125).
A function generator (HP32120A) was used to produce a modulation pattern and deliver the driving voltage waveform.
The emission spectra of the LD and LED are compared in Fig.\,\ref{Fig.3}(a).
The photocurrent from the 9-V reverse-biased LED was amplified and the real-time traces were captured by an oscilloscope (OSC, Tektronix MDO3104). 
The solid line (red) in Fig.\,\ref{Fig.3}(b) shows the trailing edge of the LED photocurrent while the broken line (black) is an exponential fit to it, $\exp(-t/\tau)$, which yields $\tau=36\,\mu$s. 
The leading edge (not shown) was also an exponential with $\tau=40\,\mu$s.  

To implement the STPI, a second AM was placed in front of the LD.
The driving voltage waveforms to produce on-off illumination patterns were output from a pulse pattern generator (PPG, HP8110A) via an rf amplifier.   
We used 2$^{11}$-1 pseudo-randomized binary sequence (PRBS) non-return-to-zero (NRZ) pulses. 
The complementary PPG output was used to emulate the reference arm, which renders the system semi-computational.
The temporal mask was NRZ-encoded using the second PPG as "101100101100" where 1 (0) corresponds to high (low) transmittance. 
The word length was 100\,$\mu$s (=\,120\,kbps) comparable with the $\tau$ of the LED.
The LED photocurrent was real-time monitored on the OSC as above.

Figure\,\ref{Fig.3}(c) shows the mask image taken in the test arm by averaging detector readouts over an ensemble of 10$^3$ realizations of random illumination pattern.
Such a direct capture of a realistic mask image as ground truth is necessary in STPI since a captured image can be slightly different from the original design.
In fact, the pulse burst pattern of the mask is clearly visible with some unintended changes in pulse width.
The solid line (orange) in Fig.\,\ref{Fig.3}(d) shows the as-captured STPI image, $M_{\rm GI,O}(t)$, obtained by using Eq.\,(\ref{eq2}), whereas the slowly-varying solid line (black) shows the trailing part of the \emph{time-reversed} impulse response of the LED, $\mu_{\rm LED}(T\!-\!t)$.

Apparently, the STPI trace decays backward in time, the envelope of which scales with $\mu_{\rm LED}(T-t)$.
The vertical dash-dotted line (black) drawn at $t\!=\!72\, \mu$s indicates when it is by $\tau$ earlier than the one-time readout at $T$.
It is seen that the latter half of the time mask "101100" is well retrieved, which spans almost twice the length of $\tau$.
Figure\,\ref{Fig.3}(e) shows the \emph{corrected} STPI image $M_{\rm GI,R}(t)$ made available through division by $\mu_{\rm LED}(T\!-\!t)$.
As indicated by the dotted lines, the high and low levels are clearly visible.
Shown in Fig.\,\ref{Fig.3}(f) for comparative purposes is the TGI image using a bucket detection which reads, $M_{\rm GI,B}(t)\!=\!\ev{\Delta I_{\rm R}(t)\Delta B}$ where $\Delta B\!\equiv\!B\!-\!\ev{B}$ with $B\!=\!\!\int \!dt' M(t')I_{\rm T}(t')$ and $\ev{\Delta I_{\rm R}(t)\Delta I_{\rm T}(t')}\!\!=\!\!\delta(t\!-\!t')$.

\begin{figure}[!b]
\begin{center}
\includegraphics[width =0.95 \linewidth]{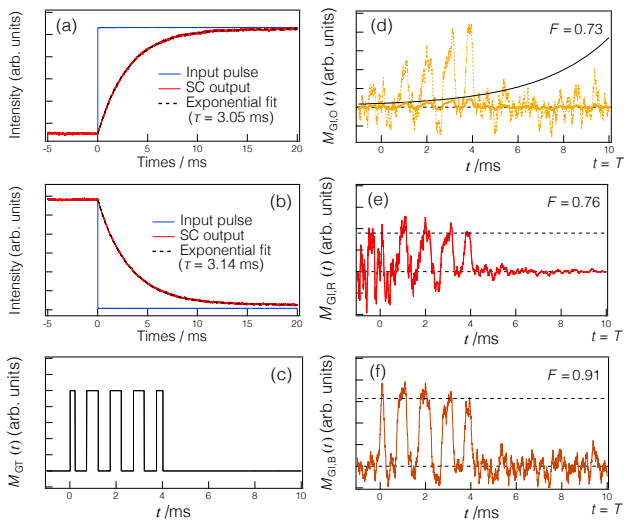}
\end{center}
\vspace{-5mm}
\caption{(a) Leading-edge and (b) trailing-edge part of response function of solar cell as photodetector, $\mu_{\rm SC}$. Broken lines are an exponential fit with $\tau\!=\!3.05$ ms (a) and $\tau\!=\!3.14$ ms (b). (c) As-captured temporal mask profile. (d) As-captured STPI image (orange) and the time-reversed $\mu_{\rm SC}$  (black). (e) Division-corrected STPI image. (f) TGI image.}
\label{Fig.4}
\end{figure}

Fidelity $F$ is used as a measure of the image quality since the \emph{shape} is our primary concern.
It is defined as\cite{o-oka2023} 
\begin{equation}
F(M_{\rm GT}, M_{\rm *})
\!=\!
\frac{\int\!\!dx\,\Delta M_{\rm GT}(x)\,\Delta M_{\rm *}(x)}{\sqrt{\int\!\!dx\,\Delta M^2_{\rm GT}(x)\int\!\!dx'\,\Delta M^2_{\rm *}(x')}}
\label{eq3}
\end{equation}
where GT stands for the ground truth, and is used piecewise over the intervals of interest. 
This is because there is a clear tendency that fluctuation is renormalized to be larger at earlier times.
For $t<T\!-\!\tau\!$, a low fidelity $F$ = 0.55 is found as opposed to $F$=0.88 for the conventional TGI. 
However, for the interval $T\!-\!\tau\!<\!t\!<\!T$, $F$=0.96 is obtained, which compares well with $F$=0.93 for the TGI.

In the second part, the STPI was implemented by using SC as the photodetector.
We used a polycrystalline Si SC (Goldmaster \& Ever Step Development Ltd, ETM250-0.5V).
Figure\,\ref{Fig.2}(b) shows the schematic setup. 
The response function of the SC, $\mu_{\rm SC}$, was measured first.
The same LD was used as the light source as in the first part. 
Programmed illumination patterns were generated by the faster PPG.
1-kbps pulse bursts at 10-Hz repetition were used to drive the AM placed at the LD output.
The voltage across a shunt resistance of the solar cell was monitored in real-time on the OSC. 
The rectangular profiles (blue) in Figs.\,\ref{Fig.4}(a) and (b) represent the input pulse while the curves (red) show the SC response. 
From the exponential fits $\exp(t/\tau)$ for (a) and $\exp(-t/\tau)$ for (b), shown by the broken lines, we obtain $\tau=3.05$ ms for (a) and $\tau=3.14$ ms for (b), respectively, which are two orders of magnitude larger than the $\tau$ of the LED.

The STPI was implemented by using 2$^{11}$-1 PRBS NRZ pulse patterns of 10-kHz clock rate to drive the AM.
An auxiliary PPG output was coupled into the OSC as the reference signal of the illumination pattern.
An optical chopper was used as the time mask in place of AM.
To produce a pulse burst, a beam blanking circuit essentially a counter-timer was inserted in between the AM and PPG, which was triggered by the reference output from the chopper driver.
Specifically, we used a 1-kbps intensity modulation within a time frame of 4 ms (=250\,bps), which is comparable with the $\tau$ of the SC. 

\begin{figure}[!b]
\begin{center}
\includegraphics[width =1 \linewidth]{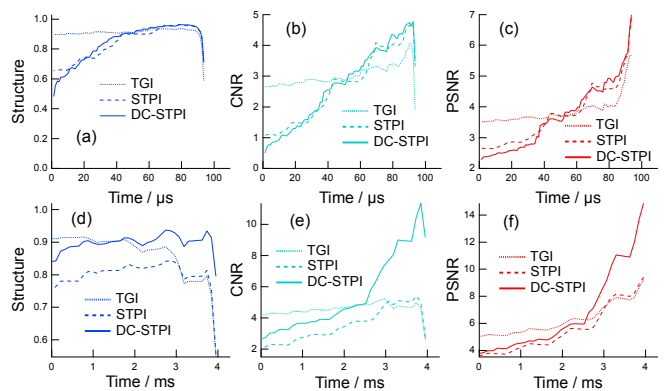}
\end{center}
\vspace{-5mm}
\caption{Image quality metrics versus time. (Upper row) LED, (Lower row) SC. (a)(d) Structure. (b)(e) Contrast-to-noise-ratio. (c)(f) Peak-signal-to-noise-ratio.}
\label{Fig.5}
\end{figure}

Figure\,\ref{Fig.4}(c) shows the mask image, $M_{\rm GT}(t)$, taken in the test arm by averaging 10$^3$ data sets. 
Figure\,\ref{Fig.4}(d) shows the as-captured STPI image, $M_{\rm GI,O}(t)$, (orange) and $\mu_{\rm SC}$  (black).
As with the LED case, the overall trace of the former scales with the latter, which permits correction by division. The result, $M_{\rm GI,R}(t)$, is shown in Fig.\,\ref{Fig.5}(e).
For comparative purposes, the TGI result, $M_{\rm GI,B}(t)$, is shown in Fig.\,\ref{Fig.4}(f).
Judging from the horizontal broken lines that discriminate the high and low levels, such division-correction was largely successful.
Notably, however, the fidelity $F\!=\!$ 0.76 calculated using the entire trace is not improved much as compared with $F\!=\!0.73$ before correction. This is in contrast with the case of the LED, and also with $F\!=\!$0.91 for the conventional TGI.
This is due to the noisy trace at at early times.
Thus the fidelity is likely to be compromised for extended operation, which raises an issue.
Roughly speaking, $M(t)$ is retrievable up to two to three times $\tau$ in time length.

Finally we use the well-known image quality metrics to evaluate the acquired images time-wise.
This is motivated by the fact that noise distribution is apparently biased due to $\mu$ with more noise at earlier times. 
In view of this, an attempt is made to single out the most appropriate one.
A time-dependent metric $A$ is such that
\begin{equation}
A(t)\!=\!\ev{A\!\in\![T\!-\!t,T]}.
\label{eq3}
\end{equation}
In the upper row of Figure\,\ref{Fig.5}, (a) structure\cite{wang2004image}, (b) contrast-to-noise ratio (CNR)\cite{zhang2018tabletop} and (c) peak-signal-to-noise ratio (PSNR)\cite{qi2023optimization} of the LED are plotted against time.  
For the SC, the same group of metrics is shown in the lower row in  lexicographic order (d)-(f).
Here the regularization parameter of the structure is c$_3$=4.5$\times 10^{-4}$.
The three metrics have been shortlisted in the sense that the as-captured STPI, and division-corrected STPI modes of operation excel the TGI at the last light input. 

The values for the STPI-relevant modes decay, if not monotonically, towards the time origin.
The division-corrected STPI $M_{\rm GI,R}(t)$, is largely better than the as-captured STPI, $M_{\rm GI,O}(t)$.
As to the LED, however, both fall below $M_{\rm GI,B}(t)$ with respect to the structure and CNR for $t\!<\!40\,\mu$s.
Thus it is appropriate to discuss the duration $\tau_{\rm eff}$ during which our STPI works effectively by referring to where the metric curves cross. 
One finds that $\tau_{\rm eff}\!=\!2\!\--\!2.5\,\tau$ for the LED, whereas $\tau_{\rm eff}\!\gtrsim 3\,\tau$ for the SC.
In view of these, the $\tau_{\rm eff}$ is likely to depend on the specific choice of a pulse sequence. 
Redesigning the the illumination pattern might bring an improvement in these metrics as partitioning did in the previous studies\cite{tang2018single,Xinwei_CTGI_LED} . 
In this regard, our STPI merits further dedicated study. 

\section*{Acknowledgements}
\noindent
This work was in part supported by JSPS KAKENHI JP21H05585. 

\bibliography{STPI_1_2.bib}

\end{document}